\documentclass[aps,prd,twocolumn,groupedaddress,superscriptaddress,floatfix,showpacs,showkeys]{revtex4}

\usepackage[utf8]{inputenc}
\usepackage[T1]{fontenc}
\usepackage{slashed}
\usepackage{times}

\usepackage{amssymb,amsfonts,amsmath,amsthm}
\usepackage{dsfont,bbm}
\usepackage{dcolumn}

\usepackage{graphicx}
\usepackage[caption=false]{subfig}

\begin{document}

\title{Single spin
asymmetries in forward p-p/A collisions revisited: the role of color entanglement}

\author{Jian~Zhou}
\affiliation{School of Physics and Key Laboratory of Particle Physics and Particle Irradiation
(MOE),  Shandong University, Jinan, Shandong 250100, China}

\begin{abstract}
\noindent We calculate the  single transverse spin asymmetries(SSA) for forward inclusive particle
production in  pp and pA collisions using a hybrid approach. It is shown that the Sivers type
contribution to the SSA drops out due to  color entanglement effect, whereas the fragmentation
contribution to the spin asymmetry is not affected by color entanglement effect. This finding
offers a natural solution for the sign mismatch problem.
\end{abstract}

 \maketitle
\noindent {\it I. Introduction.}\,---\, During the past three decades, the studies of transverse
single spin asymmetries in high energy scatterings have greatly deepened our understanding of some
central aspects of Quantum Chromodynamics(QCD) factorization theorem, among which the universality
issue attracted a lot of attentions. Within transverse momentum dependent(TMD) factorization
framework~\cite{Collins:1981uk}, a TMD distribution, known as the Sivers function
$f_{1T}^{\perp}$~\cite{Sivers:1989cc} was proposed to account for the observed large SSAs. It has
been found that the Sivers function reverses sign between the semi-inclusive deeply inelastic
scattering(SIDIS) and the Drell-Yan process~\cite{Collins:1992kk,Brodsky:2002cx,Collins:2002kn}.
The discovery of such  novel and unique universality property has stimulated  a lot of theoretical
progress over the past decade. The preliminary results from the STAR collaboration and the COMPASS
collaboration~\cite{Adamczyk:2015gyk,Aghasyan:2017jop} seem to confirm the sign change. This is
undoubtedly one of the most remarkable achievements in high energy spin physics.

However, the situation with the SSA for the forward inclusive hadron production in $pp$
collisions(denoted as $A_N$) $p^\uparrow p\rightarrow h X$ is more complicated. Due to the lack of
an additional hard scale, it is more appropriate to compute this observable using the collinear
twist-3 approach~\cite{Efremov:1981sh,Qiu:1991pp,Kouvaris:2006zy,Yuan:2009dw,Metz:2012ct} instead
of TMD factorization. Phenomenologically, it was also studied in the generalized parton
model~\cite{Anselmino:1994tv,Anselmino:2005sh}. The twist-3 effects leading to the SSA can be
factorized into various three-parton correlation functions. One of these is the Qiu-Sterman
function $T_F$~\cite{Qiu:1991pp} which can be related to the Sivers function~\cite{Boer:2003cm},
\begin{equation}
T_F(x,x)=-\!\int \! d^2 p_\perp  \frac{p_\perp^2}{M} f_{1T}^{\perp}(x,p_\perp^2)|_\text{SIDIS}
\end{equation}
where $M$ is the nucleon mass. Due to this relation, one can determine $T_F$ using the date on the
SSA measured in SIDIS and compared with the Qiu-Sterman function extracted from inclusive hadron
production in $pp$ collisions. Very surprisingly,  $T_F$ extracted from these two observables
actually differ in sign~\cite{Kang:2011hk}. To resolve this sign mismatch problem, the authors of
Ref.~\cite{Kanazawa:2014dca} suggested that a genuine twist-3 function $ \text{Im} \hat E_{\text
F}$~\cite{Yuan:2009dw}($\hat H_{\text{FU}}^{\Im}$ in a different notation) instead of $T_F$ gives
rise to the dominant contribution to $A_N$. It is worthy to mention that the data on the SSA in
SIDIS~\cite{Airapetian:2009ab,Katich:2013atq} doesn't disfavor this point of view because the
Sivers function is not well constrained at large $x$ in SIDIS, allowing  flexible parametrizations
of $T_F$. Note that all other possible sources contributing to $A_N$ in the collinear twist-3
approach were shown to be small~\cite{Kanazawa:2000hz,Koike:2007rq,Kanazawa:2010au}.

 The study of the SSA for inclusive hadron production in $pA$ collisions
$p^\uparrow A\rightarrow h X$ could play an important role  in pining down the true main cause of
$A_N$ since the different sources contributing to $A_N$ is affected by saturation effect in the
different ways. In fact, no strong nuclear suppression was observed in a recent measurement of
$A_N$ in forward $pA$ collisions~\cite{Heppelmann:2016siw}. This implies that the dominant pieces
must be these which are not affected by saturation effect. It is thus of a great interest to take
into account the saturation effect on the unpolarized target side. Some earlier work in this
direction have been done in Refs.~\cite{Boer:2006rj,Kang:2011ni,Kovchegov:2012ga,Kang:2012vm}.

In this paper,  we compute $A_N$  using a hybrid approach~\cite{Schafer:2014zea,Zhou:2015ima} where
target nucleus(or proton) is treated in the color glass condensate(CGC)
framework~\cite{McLerran:1993ni}, while collinear twist-3 approach is applied to the transversely
polarized projectile. It is a natural and powerful approach to take into account color entanglement
effect that was first discovered in Ref.~\cite{Rogers:2010dm}(for the relevant work, see
Refs.~\cite{Buffing:2011mj,Buffing:2012sz,Rogers:2013zha}). Actually, it has been found that the
SSAs for prompt photon production and photon-jet production in $pp$ or $pA$ collisions receive the
contribution from color entanglement effect~\cite{Schafer:2014zea,Schafer:2014xpa}. In contrast,
color entanglement effect is absent in the Drell-Yan process at low transverse momentum due to the
trivial color flow~\cite{Zhou:2015ima}.

We notice that the hybrid approach has been used to compute $A_N$ in
Refs.~\cite{Hatta:2016wjz,Hatta:2016khv} in the dilute limit. To include saturation effect, the
authors of Ref.~\cite{Hatta:2016wjz} derived the Wilson line structure using some heuristic
argument, which, however differs from that we directly derived  in the hybrid approach for the
Sivers type contribution. To be more explicit, the Wilson line structure we obtained can be cast
into the combination $G_{\text {DP}}- N_c^2 G_4$ where $G_{\text {DP}}$ is the normal dipole type
gluon distribution, and $G_4$ is the gluon distribution that arises from color entanglement effect.
Quit dramatically, the relation $G_{\text {DP}}= N_c^2 G_4$ holds in a quasi-classical model
indicates that the Sivers type contribution completely drops out. As explained below, the heuristic
argument used in Ref.~\cite{Hatta:2016khv} to work out the Wilson line structure in the
fragmentation case is well justified.  It is shown~\cite{Hatta:2016khv} that the contribution from
the twist-3 fragmentation function related to the moment of the TMD Collins
function~\cite{Collins:1992kk,Yuan:2009dw} is strongly suppressed by saturation effect.  In view of
the recent measurement at RHIC~\cite{Heppelmann:2016siw}, the genuine twist-3 fragmentation
function turns out to be the only candidate for the main cause of $A_N$.

\noindent
{\it II. The computation of $A_N$ in the hybrid approach}\,---\, We start the computation
of $A_N$ in the hybrid approach by introducing the relevant kinematics. The dominant partonic
channel for the spin independent forward particle production is,
\begin{equation}
q_p(xP)+ g_A(x_g' \bar P+k_\perp) \to q(l_q)
\end{equation}
which represents a quark $q_p$ from proton scattering off  classical background gluon field $g_A$
inside the target. The light cone momenta are defined as $\bar P^\mu=\bar P^- n^\mu$ and $ P^\mu=
P^+ p^\mu$ with the usual light cone vectors $n^\mu$ and $p^\mu$, normalized according to $p \cdot
n=1$.

To generate an imaginary phase necessary for the nonvanishing spin asymmetry, one additional gluon
attachment from the remanent of the polarized proton projectile must be taken into account. It is
convenient to formulate such twist-3 calculation in the covariant gauge in which this extra gluon
is longitudinally polarized. One then has to sum the multiple re-scattering of the incoming quark
and the collinear gluon with small $x$ gluon field inside  target to all orders
 simultaneously.

The incoming quark and gluon with physical polarization scattering off  CGC state can be summed
into a Wilson line  in the  fundamental and adjoint representation, respectively,
\begin{eqnarray}
 U(x_\perp)&=& {\cal P} {\rm exp} \left [ ig \int_{-\infty}^{+\infty} dz^+ A_A^-(z^+,x_\perp) \cdot t \right ]
 \\
 \tilde U(x_\perp)&=&{\cal P} {\rm exp} \left [ ig \int_{-\infty}^{+\infty} dz^+ A_A^-(z^+,x_\perp)
\cdot T \right ]
\end{eqnarray}
with $T$  and $t$ being the generators in the adjoint and fundamental representation.
 However, the multiple scattering of a longitudinally polarized gluon with
the background gluon field of target can not be simply described by a Wilson line in the CGC
formalism.  Instead, the expression for the gauge field created through the fusion of
longitudinally polarized gluon from the proton and small $x$ gluons from the target takes a quite
complicate form~\cite{Blaizot:2004wu}. It contains both singular terms(proportional to
$\delta(z^+)$)  and the regular terms: $ A^\mu=A^\mu_{reg}+\delta^{\mu-} A^-_{sing} $, whose
explicit expressions  can be found in Refs.~\cite{Blaizot:2004wu}.
\begin{figure}[t]
\begin{center}
\includegraphics[width=9 cm]{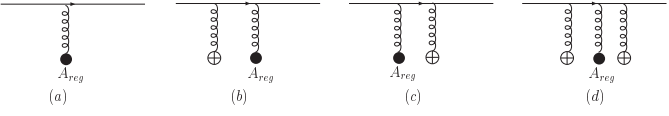}
\caption[] { The  contribution  from the regular terms to the spin dependent amplitude.  A black
dot denotes a classical field $A_{reg}$ insertion.}
 \label{1}
\end{center}
\end{figure}
\begin{figure}[t]
\begin{center}
\includegraphics[width=9 cm]{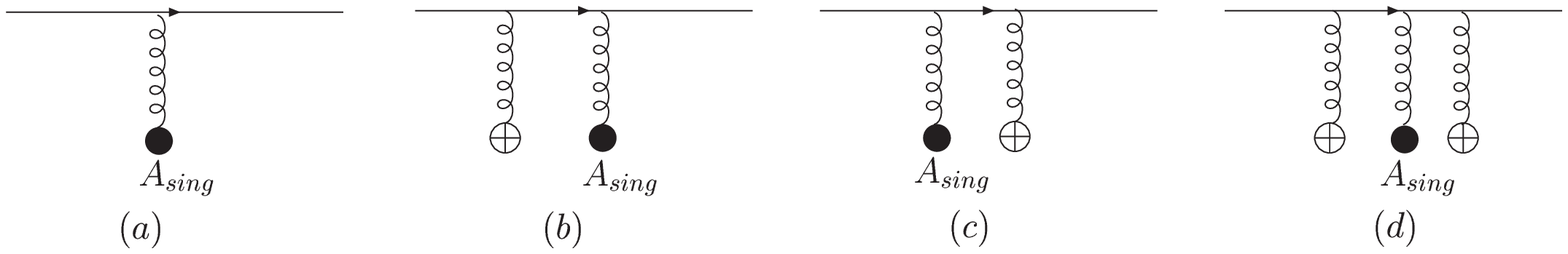}
\caption[] { The  contribution from the singular terms to the spin dependent amplitude.  A black
dot denotes a classical field $A_{sing}$ insertion. }
 \label{1}
\end{center}
\end{figure}

When computing the spin dependent  amplitude, all possible insertions of the fields $A^\mu_{reg}$
and $A^-_{sing} $ on the quark line must be taken into account as illustrated in Fig.1 and Fig.2
respectively. We calculate the contributions from Fig.1 and Fig.2
 following the method outlined in Refs.~\cite{Schafer:2014zea,Zhou:2015ima}. Note that  Fig.1(c) and Fig.1(d)
don't contribution to the  amplitude because two poles are lying on the same half plane. The final
expression for the spin dependent amplitude takes form,
\begin{eqnarray}
{\cal M}\!\!&=&\!\!\!- g \!\!  \int \!\! \frac{d k^-_1 d^2 k_{1\perp} d^2 x_\perp d^2 x_{1\perp}
}{(2\pi)^3}
 e^{ix_\perp \! \cdot (k_\perp\! -k_{1\perp}\! -p_\perp)}e^{ix_{1\perp} \! \cdot k_{1\perp}}
\nonumber \\
&& \! \!\! \!\! \! \! \!\! \!\! \!  \times  \bar u(l_q)\!
  \frac{C_U \!\!\!\!\!\!\!/  \ \ (q,p_\perp)}{q^2+i\epsilon}  t^b S_{\!F}(l_q\!-\!q) n\!\!\!/U\!(x_{\perp})
 u(xP) \!\! \left [ \tilde U\!(x_{1\perp})\!-\!1 \right ]_{ba}
  \nonumber \\
&& \!\! \!\! \!\! \!\! \!\!\!\! \!\! \!+ g  \!\! \int \!\! d^2 x_\perp e^{i(k_\perp\!-p_\perp)
\cdot x_\perp} \! \frac{\bar u(l_q) n \!\!\!/ t^b U\!( x_\perp)  u(xP)} {x_gP^++i\epsilon}\! \!
\left [ \tilde U\!(x_\perp)\! -\!1 \right ]_{ba} \nonumber \\ && \!\! \!\! \!\! \!\! \!\!\!\!
\!\!\! + ig \!\! \int \!\! d^2 x_\perp e^{i(k_\perp\! -p_\perp)  \cdot x_\perp} \! \bar u(l_q) t^a
p\!\!\!/ S_F(l_q\!\!-\!x_gP\!-\!p_\perp) n\!\!\!/ u(xP)
  \nonumber \\ &&  \! \!\! \!\! \! \! \!  \times
\left [ U(x_{\perp})-1 \right ]
   \label{pamp}
\end{eqnarray}
where  the color index $a$ is associated with the collinear gluon from the polarized projectile
which carries momentum $x_g P+p_\perp$.  $S_F(l_q-q)$ and $S_F(l_q-x_gP-p_\perp)$ denote the
standard quark propagators. The four vector $C_U^\mu(q,p_\perp)$ is defined as,
\begin{eqnarray}
&& C^+_U(q,p_\perp)=-\frac{{ p}_\perp^2}{q^-+i\epsilon}, \ \
C_U^-(q,p_\perp)=\frac{{k}_{1\perp}^2-{q}_\perp^2}{q^++i\epsilon},
 \\ \nonumber
 &&
C^i_U(q,p_\perp)=-2{\bold p}_\perp^i \label{twist2}
\end{eqnarray}
where $q^\mu=x_{g1}'\bar P^\mu+k_{1\perp}^\mu+x_g P^\mu+p_\perp^\mu$. The notation ${\bold
p}_{\perp}$ is used to denote four dimension vector with  $p_{\perp}^2=-{\bold p}_\perp^2$.

It is worthy to point out that the second term in Eq.~\ref{pamp} which describes the interaction
between the collinear gluon from the projectile and color source inside the target is missing in
Ref.~\cite{Hatta:2016wjz}, and  the Wilson line structure in the rest two terms are also organized
in different ways as compared to that in Ref.~\cite{Hatta:2016wjz}. Before computing the twist-3
piece, as a consistency check, let us first have a look at the twist-2 part of the derived
amplitude by setting $p_\perp=0$.  The first term vanishes due to $C_U^\mu(q,p_\perp=0)=0$. The
leading twist contribution of the amplitude is simplified as,
\begin{eqnarray}
{\cal M}_{\text{twist-2}} \!\!   &=& \!\! \frac{g}{P^+}  \!\! \int \!\! d^2 x_\perp e^{ik_\perp
\cdot x_\perp}  \nonumber
\\ && \!\! \!\! \!\! \!\! \!\!\!\! \!\!\! \left \{ \left[ {\cal P}\frac{1}{x_g}+i \pi \delta(x_g) \right ]
 \left [ U(x_{\perp})-1 \right  ] \bar u(l_q) n \!\!\!/ t^a u(xP)  \right .\
 \nonumber
\\ && \!\! \!\! \!\! \!\! \!\!\!\!  \left .\  -i \pi \delta(x_g)
  \bar u(l_q) n \!\!\!/ t^a u(xP) 2 \left [ U(x_{\perp})-1 \right  ] \right \}
   \label{twist2}
\end{eqnarray}
In arriving at the above expression, we used the algebraic identity, $U(x_{\perp})t^b
U^\dag(x_{\perp})\!=\! t^a \tilde U_{ba}(x_\perp) $. After integrating out the incoming quark
transverse momentum, the contributions proportional to the delta function $\delta(x_g)$ are
canceled out between the different cut diagrams. The additional gluon exchange from proton can be
incorporated into the gauge link appears in the matrix element definition of quark PDF by carrying
out the $x_g$ integration over the principal value part.  As expected, the corresponding hard part
is just the Born diagram contribution to a quark scattering off CGC state~\cite{Dumitru:2002qt}. At
this point, one can readily see that it is critical to keep the scattering amplitude gauge
invariant by taking into account the initial interaction with the color source inside target. Note
that the result derived in Ref.~\cite{Hatta:2016wjz} fails to pass this consistency check.

If one applies TMD factorization on the polarized projectile side,  the terms proportional to the
delta function contributes to the gauge link in the Sivers TMD function.  But unlike photon-jet
production~\cite{Schafer:2014xpa}, such hybrid approach might not be well justified in the process
under consideration because of the lack of an additional hard scale.

We now proceed to compute the spin dependent twist-3 contribution by first isolating imaginary part
from different poles. We start with analyzing the pole structure in the first term  in
Eq.~\ref{pamp}. By carrying out $x_g$ and $k_1^-$ integration, two propagators are effectively put
on shell,
\begin{eqnarray}
q^2=0 \ , \ \ (l_q-q)^2=0
\end{eqnarray}
Three particle lines connected by a quark-gluon vertex being  simultaneously on shell  implies that
three momenta $q^\mu, l_q^\mu-q^\mu, l_q^\mu$ must be collinear to each other. This leads to,
\begin{eqnarray}
\bar u(l_q) C_U \!\!\!\!\!\!\!/  \ \ (q,p_\perp) (l_q \!\!\!\!\!/-q\!\!\!/)=-\bar u(l_q) l_q
\!\!\!\!\!/ \ \ C_U \!\!\!\!\!\!\!/  \ \ (q,p_\perp) (1-\beta)=0
\end{eqnarray}
where $q^\mu=\beta l_q^\mu$ for $0\leq\beta \leq1$. When commuting $C_U \!\!\!\!\!\!\!/$ \ \  with
$l_q \!\!\!\!\!/-q\!\!\!/$ in the above formula, we used the property $C_U^\mu(q,p_\perp) \cdot
q_\mu=0$. One thus concludes that the hard gluon pole(or the soft fermion pole for $\beta=1$)
contribution is completely washed out by saturation effect.  This analysis is in agreement with
that made in Ref.~\cite{Hatta:2016wjz}.

One should notice that the first term in Eq.~\ref{pamp} also contains the soft gluon pole(SGP)
contribution which comes from the minus component of $C_U^\mu$. Combining it with the last two
terms in Eq.~\ref{pamp}, the SGP contribution is given by,
\begin{eqnarray}
{\cal M}_{\text {SGP}}\!\!&=&\!\!\!-i\pi g \!\!  \int \!\! \frac{ d^2 k_{1\perp} d^2 x_\perp d^2
x_{1\perp} }{(2\pi)^2}
 e^{ix_\perp \! \cdot (k_\perp\! -k_{1\perp}\! -p_\perp)}e^{ix_{1\perp} \! \cdot k_{1\perp}}
\nonumber \\
&&\times \delta(x_g\!P^+) \frac{k_{1\perp}^2}{q_\perp^2} \ \bar u(l_q) n \!\!\!/ t^b U\!( x_\perp)
u(xP)  \tilde U\!(x_{1\perp})_{ba}\!\!
 \nonumber \\ && \!\! \!\! \!\!\!
 + i\pi g \!\! \int \!\! d^2 x_\perp e^{i(k_\perp\! -p_\perp)  \cdot x_\perp}
\!\delta((l_q-x_gP-p_\perp)^2)
  \nonumber \\ &&    \times
   \bar u(l_q) t^a
p\!\!\!/ ( l_q\!\!\!\!\!/-\!x_g\!P\!\!\!\!/-\!p_\perp\!\!\!\!\!\!\!/ \ \ ) n\!\!\!/ u(xP) \left [
U(x_{\perp})-1 \right ]
   \label{sfp}
\end{eqnarray}
where the last term gives rise to the so-called derivative term contribution.  At this point, we
would like to mention that the spin dependent amplitude takes a slightly different form for the
left cut diagrams due to the different $p_\perp$ flow.  In the collinear twist-3 approach, the spin
asymmetry arises from the interference between the imaginary part identified above and the
conjugate Born scattering amplitude without an additional gluon attachment from the projectile. It
is straightforward to compute the later in the CGC formalism~\cite{Dumitru:2002qt}. Following the
standard procedure, the next step is to make $p_\perp$ expansion, and factorize the soft part on
the polarized proton side into the Qiu-Sterman function.

Finally, in order to express the spin dependent cross section in terms of the known gluon
distributions, we simplify the relevant color structure, starting with the one associated with the
delta function $\delta(x_g P^+)$,
\begin{eqnarray}
&&\!\!\!\!\! \!\!\!\!\! {\rm Tr} \! \left [ t^a U^\dag(y_\perp) t^b U(x_\perp) \right ] \! \tilde
U(x_{1\perp})_{ba}\! = \! \frac{-1}{2N_c} {\rm Tr} \! \left [U^\dag(y_\perp) U(x_\perp) \right
]\nonumber \\ && \ \ \ \ \ \ \ \ \ \ \ \  +\frac{1}{2} {\rm Tr}\! \left[
U^\dag(y_\perp)U(x_{1\perp}) \right ] {\rm Tr} \left [ U^\dag(x_{1\perp})U(x_\perp) \right ]
\end{eqnarray}
where $ U^\dag(y_\perp) $  is from the conjugate amplitude. Note that the forward scattering
amplitude contribution has been neglected as we do so below. The contribution from $\left [
U^\dag(y_\perp) U(x_\perp) \right ]$ drops out because one can trivially carry out $x_{1\perp}$
integration, resulting in $k_{1\perp}=0$. In the large $N_c$ approximation, $\langle {\rm Tr}\!
\left[ U^\dag(y_\perp)U(x_{1\perp}) \right ] {\rm Tr} \left [ U^\dag(x_{1\perp})U(x_\perp) \right ]
\rangle$  can be related to the convolution of two dipole type gluon distributions. After summing
the left and right cut diagrams contribution and making $p_\perp$ expansion, we encounter the
following structure,
\begin{eqnarray}
 \int \!\! d^2 k_{1\perp} \!\! \left [
 \frac{l_{q\perp}^\alpha-k_{1\perp}^\alpha}{(l_{q\perp}-k_{1\perp})^2}
 F(l_{q\perp}^2)+ \frac{l_{q\perp}^\alpha}{2}
 \frac{\partial F(l_{q\perp}^2)}{\partial l_{q\perp}^2} \right ] \!
 F(k_{1\perp}^2)
\end{eqnarray}
where $F(l_{q\perp}^2)$ is the Fourier transform of the dipole amplitude whose definition is given
below. Using the method introduced in Ref.~\cite{Hatta:2016khv}, it is easy to verify that two
terms are completely canceled out in the dilute limit, and are strongly suppressed in the
saturation regime. One thus can safely neglect the SGP contribution induced by the initial state
interaction.

We now turn to discuss the Wilson lines associated with the derivative term contribution, which
reads,
\begin{eqnarray}
 {\rm Tr} \! \left [ t^a U^\dag(y_\perp) t^a U(x_\perp) \right ] \!\!  &=& \!\!
\frac{1}{2} {\rm Tr}\! \left[ U^\dag(y_\perp) \right ] {\rm Tr}\!  \left [U(x_{\perp}) \right ]
\nonumber \\ &-& \!\! \frac{1}{2N_c} {\rm Tr} \left [ U^\dag(y_{\perp})U(x_\perp) \right ]
\end{eqnarray}
where the  non-trivial color  structure ${\rm Tr}\! \left[ U^\dag(y_\perp) \right ] {\rm Tr}\!
\left [U(x_{\perp}) \right ]$ arises from color entanglement effect as explained in
Refs.~\cite{Schafer:2014zea,Schafer:2014xpa,Zhou:2015ima}. The extra gluon attachment from the
polarized proton plays a crucial role in yielding such unique structure. With all these calculation
recipes, we derive the spin dependent partonic cross section,
\begin{eqnarray}
&& \!\!\!\!\!\!\!\!\!\!\!\!
 \frac{d \sigma}{dy d^2l_{q\perp}}=\frac{2\pi^2 \alpha_s x x_g'}{ N_c(N_c^2-1)}
 \frac{\epsilon_{\alpha \beta} S_{\perp }^{\beta}l_{q\perp}^{ \alpha}}{l_{q\perp}^2}
  \nonumber \\&& \!\!\!\! \times \! \left \{\! \frac{1}{l_{q\perp}^2} \left [ G_{\text{DP}}(x_g',l_{q\perp}^2)
  \!-\!N_c^2 G_{4}(x_g',l_{q\perp}^2)\right  ] x \frac{d T_F(x,x)}{d x} \right.\
\nonumber \\
&& \!\!\! \left .\ \ \ \ \  \ +\frac{\partial\left [ G_{\text {DP}}(x_g',l_{q\perp}^2)\!-\!N_c^2
G_{4}(x_g',l_{q\perp}^2)\right  ]}{\partial l_{q\perp}^2 } T_F(x,x) \!\! \right \} \label{cs}
\end{eqnarray}
where $S_\perp$ is the transverse spin vector of the proton. The momentum fractions $x$ and $x_g'$
are fixed according to $x=e^y|l_{q\perp}|/\sqrt{s} $ and $x_g'=e^{-y}|l_{q\perp}|/\sqrt{s} $ with
$y$ being the outgoing quark rapidity.  $G_{\text {DP}}$ is the normal dipole type gluon
distribution, and related to the Fourier transform of the dipole amplitude $x_g'G_{\text
{DP}}(x_g',l_{q\perp}^2)=\frac{l_{q\perp}^2N_c}{2 \pi^2 \alpha_s} F(l_{q\perp}^2)$. $G_4$
introduced in Ref.~\cite{Schafer:2014zea} is the gluon distribution that arises from color
entanglement effect. Their operator definition are given by,
\begin{eqnarray}
 x_g' G_{\text {DP}}(x_g',l_{q\perp}^2)\!\! &= & \!\! \frac{l_{q\perp}^2N_c}{2 \pi^2 \alpha_s}
 \int \! \frac{d^2 x_\perp d^2 y_\perp}{(2 \pi)^2 } e^{il_{q\perp} \! \cdot
(x_\perp-y_\perp)} \nonumber \\ && \times \frac{1}{N_c} \langle {\rm Tr} \left [
U^\dag(y_\perp)U(x_\perp) \right ]\rangle \nonumber
\\   x_g' G_{4}(x_g',l_{q\perp}^2)\!\!&= & \!\! \frac{l_{q\perp}^2N_c}{2 \pi^2 \alpha_s}
 \int \! \frac{d^2 x_\perp d^2 y_\perp}{(2 \pi)^2 } e^{il_{q\perp} \! \cdot
(x_\perp-y_\perp)} \nonumber \\ && \times \frac{1}{N_c^2}\langle {\rm Tr} \left [
U^\dag(y_\perp)\right ]  {\rm Tr} \left [ U(x_\perp) \right ]\rangle
\end{eqnarray}
which can be evaluated and related to each other in the MV model~\cite{Schafer:2014zea},
\begin{eqnarray} x_g' G_{4}(x_g',
l_{q\perp}^2)\! = \! \frac{1}{N_c^2}x_g'G_{\text {DP}}(x_g', l_{q\perp}^2) \label{mv}
\end{eqnarray}
This simple relation leads to a complete cancelation between the contributions from $G_{\text
{DP}}$ and $G_4$ in Eq.\ref{cs}. Therefore,  the Sivers type contribution to $A_N$ drops out.
Obviously, this conclusion remains true after promoting the partonic spin dependent cross section
 to the  hardron production cross section.

We now comment on the twist-3 fragmentation function contribution to $A_N$. The derivative term
contribution to $A_N$ in pp collisions was first computed in Ref.~\cite{Yuan:2009dw} in the purely
collinear twist-3 approach. The complete result was obtained in Ref.~\cite{Metz:2012ct}(see recent
reviews Refs.~\cite{Metz:2016swz,Pitonyak:2016hqh}). In order to take into account multiple gluon
rescattering effect on target side, the similar hybrid approach also can be applied in the
fragmentation case~\cite{Hatta:2016khv}. As well known, it is highly nontrivial to compute the SGP
contribution in the light cone gauge~\cite{Zhou:2009jm,Zhou:2010ui}. Since the SGP contribution
vanishes for the twist-3 fragmentation contribution~\cite{Meissner:2008yf,Gamberg:2010uw}, it is
more convenient to carry out the calculation in the light cone gauge where the additional gluon
exchange from the twist-3 fragmentation function is physically polarized~\cite{Hatta:2016khv}. A
gluon with physical polarization scattering off the background gluon field can be summarized into a
normal Wilson line in the adjoint representation. In this sense, the derivation of the Wilson line
structure  in Ref.~\cite{Hatta:2016khv} is well justified.  If one formulates such calculation  in
the covariant gauge, the fact that an imaginary phase from the scattering amplitude is not required
in the twist-3 fragmentation case would make an essential difference in deriving the color
structure. However, the detailed investigation  is beyond the scope of the current work.

We close this section with few further
remarks:\\
(1) Following the standard procedure, one can derive the BK type evolution equation for the gluon
distribution $G_4$, which will be presented in a separate publication. In the large $N_c$ limit,
the relation Eq.\ref{mv}  holds under small $x$ evolution.\\
(2) The relation Eq.\ref{mv} is a model dependent result. In general case, an incomplete
cancelation between two gluon distributions leaves some room for having tiny spin asymmetry for
inclusive jet production in pp or pA collisions~\cite{Nogach:2012sh}.
\\
(3) If the $G_4$ contribution is neglected, Eq.\ref{cs} is consistent with the collinear twist-3
result~\cite{Kouvaris:2006zy} in the dilute limit.
\\
(4) Color entanglement effect is a leading power effect and should be taken into account in the
genuine collinear twist-3 approach as well.  We plan to redo calculation in the purely collinear
framework by going beyond one gluon exchange approximation on target side.
\\
(5) T-even objects like the unpolarized twist-2 amplitude, are not affected by color entanglement
effect. The observed color entanglement effect is the consequence of the non-trivial interplay
among T-odd effect,  multiple gluon re-scattering, and the non-Abelin feature of
QCD~\cite{Rogers:2010dm,Schafer:2014zea,Schafer:2014xpa,Zhou:2015ima}.

\noindent {\it III. Summary.}\,---\, Let us now summarize the recent progress on the topic
addressed in this Letter. The sign mismatch problem was first observed in Ref.~\cite{Kang:2011hk}.
To find a way out, one naturally questions the dominance of the Sivers type contribution to $A_N$.
It was indeed found that the genuine twist-3 fragmentation function could play an important role in
generating the spin asymmetry~\cite{Kanazawa:2014dca}. Later, the authors of
Ref.~\cite{Hatta:2016khv} have sorted out the piece of the contribution from the twist-3
fragmentation functions that is not suppressed by saturation effect using a hybrid approach first
developed in Refs.~\cite{Schafer:2014zea,Zhou:2015ima}. The saturation suppressed fragmentation
contribution being the major source of $A_N$ has been ruled out by the recent
measurement~\cite{Heppelmann:2016siw}. In this work, we demonstrate that the Sivers type
contribution to the spin asymmetry drops out due to color entanglement effect. The nuclear
independent part of the genuine twist-3 fragmentation contribution turns out to be the only
candidate for the main cause of $A_N$. A recent work~\cite{Gamberg:2017gle} shows that it is almost
sufficient to account for $A_N$ by taking into account this fragmentation term alone with the input
constrained by the Lorentz invariance relation~\cite{Kanazawa:2015ajw}. We thus believe that the
sign mismatch problem has been solved.

\noindent {\it \bf Acknowledgments:} This work has been supported by  the National Science
Foundation of China under Grant No. 11675093, and by the Thousand Talents Plan for Young
Professionals.

\end {document}